\documentclass[12pt]{article}

\usepackage{jheppub}
\pdfoutput=1
\usepackage{ae,aecompl}
\usepackage{amsmath,amsfonts,amssymb}
\usepackage{graphicx,color}
\usepackage{epsfig,epsf}
\usepackage{bm}

\usepackage{relsize}
\usepackage{pifont}
\usepackage{physics}

\usepackage{hyperref}
\hypersetup{
		colorlinks=true,
		linkcolor=[rgb]{0.0,0.4,0.6},
		filecolor=magenta,
		urlcolor=[rgb]{0.0,0.4,0.6},
		citecolor=[rgb]{0.5,0.2,0.3},
	}

\usepackage{url}
\usepackage{multirow}
\usepackage{slashed}
\usepackage{graphbox}
\usepackage{subfigure}

\usepackage[T1]{fontenc}
\usepackage{array,booktabs, makecell}
\usepackage[title]{appendix}
\usepackage{tikz}
\usepackage{tikz-feynman}
\tikzfeynmanset{compat=1.0.0}

\usepackage{ulem}
\usepackage{soul}

\newcommand{\appref}[1]{\hyperref[#1]{Appendix~\ref*{#1}}}

\newcommand{\dm}{\Delta M}

\newcommand{\mtch}{M_{T^\pm}}
\newcommand{\mtz}{M_{T^0}}

\newcommand{\lh}{\lambda_{h} }
\newcommand{\lt}{\lambda_{t} }

\newcommand{\lht}{\lambda_{ht} }
\newcommand{\half}{\frac{1}{2}}
\newcommand{\quart}{\frac{1}{4}}
\newcommand{\rgg}{\mu_{\gamma\gamma}}
\newcommand{\rzg}{\mu_{Z\gamma}}
\newcommand{\hsm}{h_{SM}}
\newcommand{\abi}{ab$^{-1}$}

\title{Dark clouds to silver linings over the hyperchargeless scalar triplets}

\author[a]{Priyotosh Bandyopadhyay,}
\author[a]{Snehashis Parashar,}

\affiliation[a]{Indian Institute of Technology Hyderabad, Kandi,  Sangareddy-502284, Telengana, India}

\emailAdd{bpriyo@phy.iith.ac.in} 
\emailAdd{ph20resch11006@iith.ac.in}

\preprint{ IITH-PH-0002/25}

\begin{document}

\abstract{A real scalar triplet with zero hypercharge offers a minimal non-trivial extension of the Standard Model (SM) with a charged Higgs and a possible dark matter  or custodial symmetry breaking signature. The $Z_2$-odd inert triplet model (ITM) provides a dark matter, while the non-inert Higgs triplet model (HTM) breaks the custodial symmetry, enabling rich collider signatures. Both these models also promise the viability of a first-order phase transition (FOPT). This letter revisits both models under various theoretical and current experimental constraints, revealing a trade-off between DM and FOPT viability, and explores the resulting gravitational wave signals and collider prospects.}

\maketitle
	
\section{Introduction}
	The monumental discovery of the Higgs boson at the LHC \cite{CMS:2012qbp, ATLAS:2012yve} establishes the presence of at least one scalar multiplet necessary for the Standard Model (SM), but keeps the doors open for further scalar extensions.
	A real scalar triplet with hypercharge $Y=0$ is a well-motivated, non-trivial extension with a minimal set of dimensionless couplings-- one self-coupling and one Higgs portal-- also providing the sought-after charged scalar. The $Z_2$-odd version of the extension, called the inert triplet model (ITM), is extensively studied, with motivations such as the weakly-interacting massive particle (WIMP) dark matter (DM) and its detectability \cite{Cirelli:2005uq, Araki:2010nak, Araki:2011hm, YaserAyazi:2014jby,  Chiang:2020rcv, Katayose:2021mew, Bandyopadhyay:2024plc}, stabilising the current electroweak (EW) vacuum \cite{ Khan:2016sxm, Jangid:2020qgo, Bandyopadhyay:2025ilx}, and achieving a first-order electroweak phase transition (EWPT) necessary for the EW baryogenesis \cite{Patel:2012pi,Niemi:2018asa,Niemi:2020hto, Bandyopadhyay:2021ipw}. The non-inert version of the scalar triplet, which we will simply refer to as the Higgs triplet model (HTM), participates in the electroweak symmetry breaking (EWSB), with a small vacuum expectation value (VEV) of the triplet that violates the approximate custodial symmetry of the SM at tree level, which makes it a fascinating scenario to study at colliders \cite{Chabab:2018ert,  FileviezPerez:2022lxp, Ashanujjaman:2023etj, Butterworth:2023rnw, Ashanujjaman:2024pky, Crivellin:2024uhc, Bandyopadhyay:2024gyg, Ashanujjaman:2024lnr}, also keeping the first-order phase transition (FOPT) in conversation \cite{Bell:2020gug}. The abundance of contemporary research on these models naturally calls for a detailed perusal into their parameter space, considering the constraints from the ever-evolving experimental bounds, as well as theoretical limitations. In this letter, we intend to present a comprehensive assessment of both ITM and HTM, combining their multitude of motivations, and provide conclusive remarks on their viability for a complete theory of DM, achieving an FOPT, and collider probes of the peculiar triplet nature.
	
	\section{The models}
	
	The scalar sector of both ITM and HTM involves the $SU(2)$ Higgs doublet $\Phi$ and the triplet $\mathcal{T}$ with $Y=0$, defined as:
	\begin{equation}	
		\Phi =	\begin{pmatrix}
			\phi^+ \\ \phi^0
		\end{pmatrix}, \mathcal{T} =\half \begin{pmatrix}
			T^0 & \sqrt{2}T^+ \\ \sqrt{2}T^- & -T^0 
		\end{pmatrix},
	\end{equation}
	
	with the tree-level scalar potentials being written as:
	\begin{align}
		\begin{split}
			V_{ITM} ={}& \mu_\Phi^2 \Phi^\dagger \Phi + \mu_\mathcal{T}^2 Tr(\mathcal{T}^\dagger \mathcal{T}) +\lambda_h \abs{\Phi^\dagger \Phi}^2 \\
			{}&+ \lambda_t \abs{Tr(\mathcal{T}^\dagger \mathcal{T})} + \lambda_{ht} \Phi^\dagger \Phi Tr(\mathcal{T}^\dagger \mathcal{T}),\\
			V_{HTM} ={}& V_{ITM}(\Phi,\mathcal{T}) + A_{ht} \Phi^\dagger \mathcal{T} \Phi.
		\end{split}
	\end{align}
	
	In $V_{ITM}$, a discrete $Z_2$ symmetry is assigned on $\mathcal{T}$ under which it transforms as odd. The presence of the $A_{ht}$-term in $V_{HTM}$ breaks any discrete symmetry on $\mathcal{T}$, allowing $T^0$ to obtain a non-zero VEV $v_t \lesssim$ 3 GeV that respects the $\rho$-parameter\cite{ParticleDataGroup:2022pth}. In case of ITM, only $\phi^0$ obtains a VEV $v_h$ and triggers EWSB, leading to the pure doublet SM Higgs $h_{SM}$, and degenerate triplet scalars $T^0,T^{\pm}$, as the scalar mass eigenstates \cite{Bandyopadhyay:2024plc}. The tree-level degeneracy between $\mtch$ and $\mtz$ is lifted at one-loop level, with $\dm \sim 166$ GeV \cite{Cirelli:2005uq}, rendering $T^0$ as a DM candidate. In contrast, both $v_h,v_t$ contribute to the EWSB in HTM, leading to mixing between the scalar gauge eigenstates. This yields the mass eigenstates that include a mostly doublet, SM-like Higgs $h$, and mostly triplet scalars $H^0,H^{\pm}$, with the charged and neutral mixing angles given by \cite{Butterworth:2023rnw, Bandyopadhyay:2024gyg, Ashanujjaman:2024lnr}: 
	\begin{equation}
		\tan 2\alpha_+ = \frac{v_h v_t}{v_t^2 - \frac{v_h^2}{4}}, \, \tan 2\alpha_0 = \frac{4v_h v_t(A_{ht} - 2\lht v_t)}{v_h^2(A_{ht} - 8\lh v_t) + 8\lt v_t^3}
	\end{equation}

	%the tree-level scalar mass spectrum\cite{Bandyopadhyay:2024plc}:
	%\begin{equation}
	%\mhsm^2 = 2 \lambda_h v_h^2, \,\, \mtz^2 = \mtch^2 = \frac{1}{2} \lht v_h^2 + \mu^{2}_T
	%\end{equation}
	
	% In contrast, both the doublet and triplet VEVs $v_h,v_t$ contributes to the EWSB, leading to mixing between the scalar mass eigenstates, with the neutral and charged mixing angles given by\cite{Bandyopadhyay:2024gyg, Ashanujjaman:2024lnr}:
	%\begin{equation}
	%\tan 2\alpha_+ = \frac{v_h v_t}{v_t^2 - \frac{v_h^2}{4}}, \, \tan 2\alpha_0 = \frac{4v_h v_t(A_{ht} - 2\lht v_t)}{v_h^2(A_{ht} - 8\lh v_t) + 8\lt v_t^3}
	%\end{equation}
	%%\begin{align}
	%%h ={}& \cos \alpha_0 \phi^0 + \sin\alpha_0 T^0, \nonumber \\
	%%H^0 ={}& -\sin \alpha_0 \phi^0 + \cos\alpha_0 T^0, \nonumber \\
	%%G^\pm ={}& \cos \alpha_+ \phi^\pm + \sin\alpha_+ T^\pm, \nonumber \\
	%%H^\pm ={}& -\sin \alpha_+ \phi^\pm + \cos\alpha_+ T^\pm. \label{eq:htm_tp}
	%%\end{align}
	%
	%
	%Out of the mass eigenstates emerging from this, $h$ is the mostly doublet, SM-like Higgs boson, and $H^0,H^{\pm}$ are mostly triplet scalars, carrying masses \cite{Butterworth:2023rnw}:
	%\begin{align}\label{eq:mhtm}
	%\begin{split}
	%M_{h/H^0}^2 ={}& \lh v_h^2 \left(1\pm\frac{1}{\cos 2\alpha_0}\right) \\
	%{}& + \left(\frac{A_{ht} v_h^2}{8v_t} + \lt v_t^2 \right) \left(1\mp\frac{1}{\cos 2\alpha_0}\right)
	%\end{split}\\
	%M_{H^\pm}^2 ={}& \frac{A_{ht} (v_h^2+4v_t^2)}{4 v_t} \nonumber
	%\end{align}
	An important pre-requisite for the potential demands that $[v_h, 0]$ ($[v_h, v_t]$) is the global minima for ITM (HTM), ensured by the conditions:
	\begin{equation}
		\text{(a) }\mu_\mathcal{T}^2 > 0, \text{  or  (b) } \mu_\mathcal{T}^2 < 0 \text{ and } \frac{\mu_\mathcal{T}^4}{\lambda_t} < \frac{\mu_\Phi^4}{\lambda_h} \label{eq:truemin}
	\end{equation}
	
	The condition (b) from \autoref{eq:truemin} is necessary for the 2-step FOPT \cite{Patel:2012pi, Vaskonen:2016yiu, Niemi:2020hto}, which we will discuss later. The perturbativity of the tree-level theory imposes $\abs{\lambda_i} \leq 4\pi$, while the bounded-from-below conditions for the potential requires $\lambda_{h,t} > 0$, and $\lambda_{ht} + 2\sqrt{\lh\lt} > 0.$ \cite{Khan:2016sxm}. The demand for a $\sim125$ GeV Higgs boson expects $\lh \simeq 0.13$, confining $\lht$ to the region $2.5 \lesssim \lht \leq 4\pi$, where our further calculations will dwell. 
	
	%As both $\phi^0$ and $T^0$ obtains their VEVs, the two minimisation conditions fix these bare mass terms in HTM:
	%\begin{align}
	%\mu_\Phi^2 ={}& \half \left(A_{ht} v_t - 2 \lambda_h v_h^2  - \lambda_{ht} v_t^2\right), \nonumber \\
	%\mu_T^2 ={}& \quart \frac{A_{ht} v_t^2 - 2 \lambda_{ht} v_h^2 v_t - 4 \lambda_t v_t^3}{v_t}. \label{eq:muHTM}
	%\end{align}
	%For ITM, the bare mass terms simplify into:
	%\begin{equation}
	%\mu_\Phi^2 =- \lambda_h v_h^2, \quad \mu_T^2 = M_{T}^2 - 2 \lambda_{ht} v_h^2. \label{eq:muITM}
	%\end{equation}

\section{The dark clouds}

	Apart from the aforementioned theoretical limits, additional constraints on $\lht$, and subsequently other model parameters, emerge from the precision measurements of the SM Higgs boson decays into vector bosons. The most recent and precise ATLAS measurement of the $h_{SM}\to\gamma\gamma$ signal strength reads $\mu_{\gamma\gamma} = 1.04_{-0.09}^{+0.10}$ \cite{ATLAS:2022tnm}. A somewhat less precise, combined measurement from CMS and ATLAS exists for $h_{SM}\to Z\gamma$ signal strength $\mu_{Z\gamma} = 2.2 \pm 0.7$, which stays $1.9\sigma$ away from the SM prediction \cite{ATLAS:2023yqk}. In scalar triplet models, the deviation from SM primarily involves the charged scalar and $\lht$, with $\lht <0$ implying $\rgg,\rzg > 1$ and vice-versa \cite{Chen:2013vi}, the details of which can be found in the decay width expressions in \appref{sec:app1}. The drastically different 1$\sigma$ regions of these two measurements make it impossible to satisfy both them simultaneously, but their $2\sigma$ limits have a good overlap. Subsequently, we can assess the viability of the models in the $M_{H/T^{\pm}} - \lht$ plane, overlaying the $\rgg,\rzg$ exclusions at $2\sigma$, and other relevant bounds from various direct and indirect searches. Exclusions from the false vacuum condition in \autoref{eq:truemin} are also imposed.
	
	\begin{figure}[htbp]
		\centering
		\subfigure[]{\includegraphics[width=0.48\linewidth]{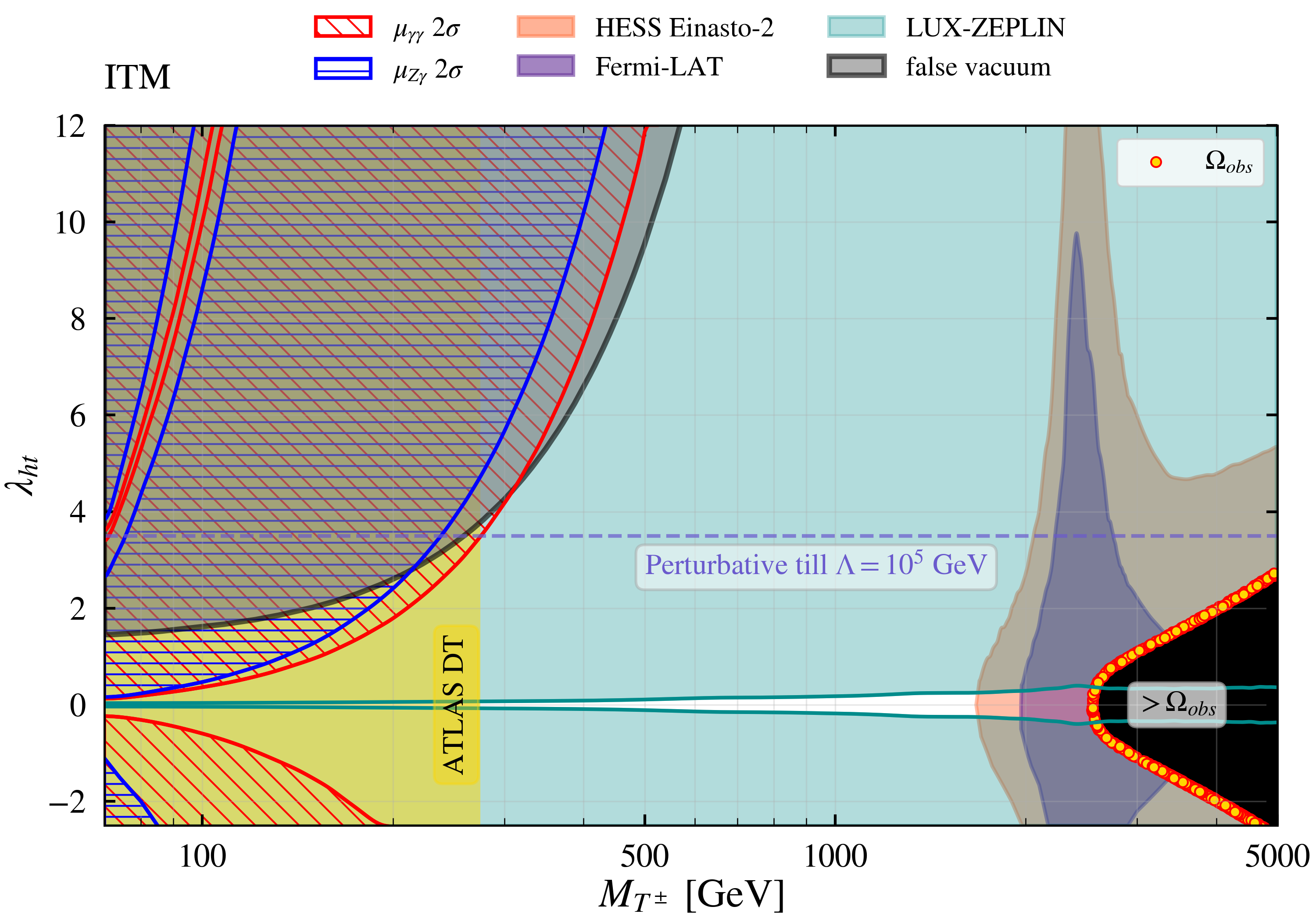}}
		\subfigure[]{\includegraphics[width=0.48\linewidth]{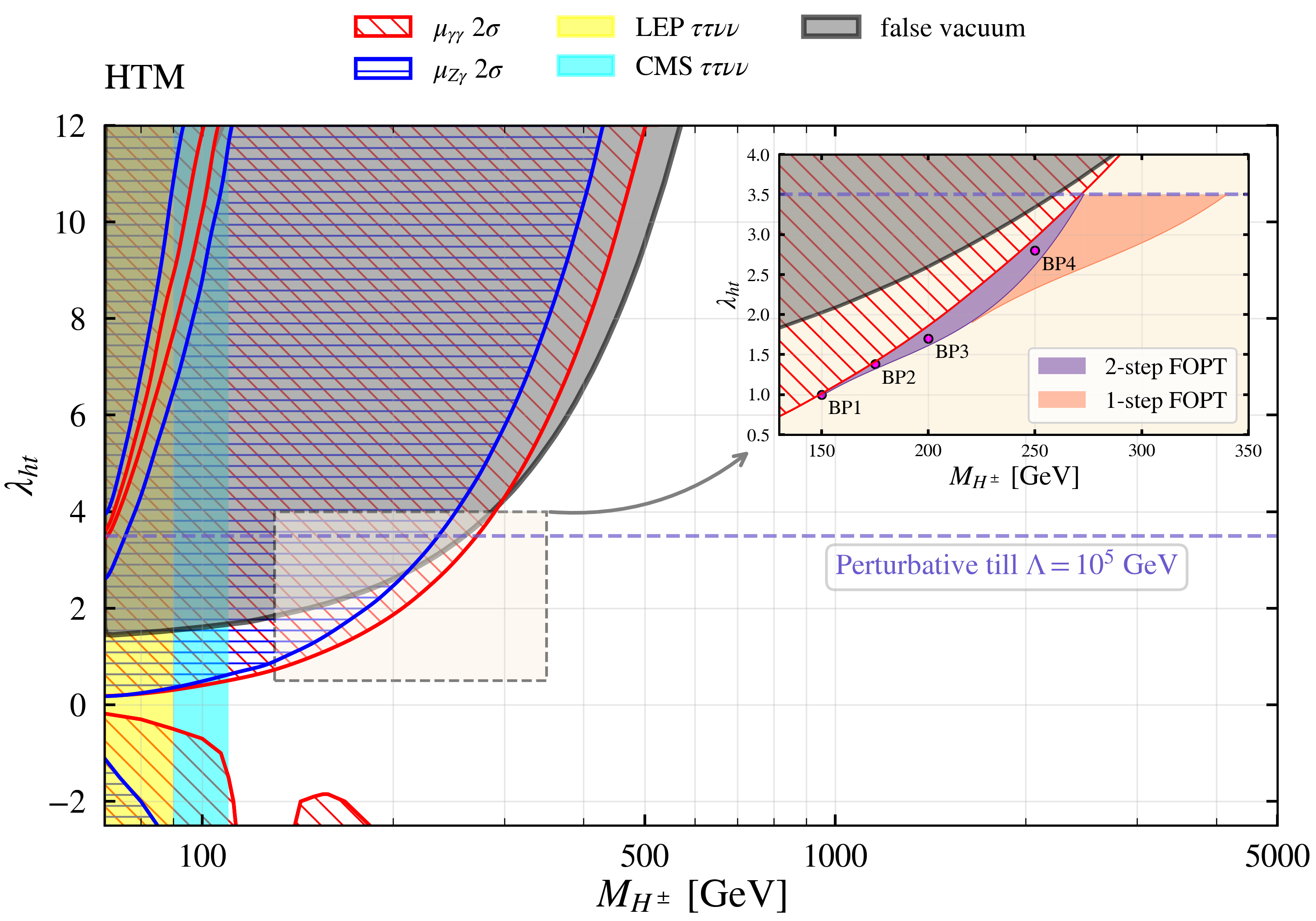}}
		\caption{Exclusion plots in the $M_{T^\pm/T^\pm} - \lambda_{ht}$ plane in case of (a) ITM, (b) HTM, for points that satisfy the $\rgg,\rzg$ measurements at $2\sigma$, with further bounds from other experiments and the false vacuum conditions.}
		\label{fig:exclusion}
	\end{figure}
	
	In \autoref{fig:exclusion}(a) we present the applicable bounds on the ITM, with $M_{T^\pm} \leq 5$ TeV. The $\rgg$ and $\rzg$ measurements at $2\sigma$ excludes the spaces covered in the red (horizontal hatch) and dark blue (diagonal hatch) regions respectively, below $M_{T^\pm} \lesssim 500$ GeV. A thin, funnel-like region within $M_{T^\pm} \in [70,110]$ GeV with $\lht \geq 3$ remains allowed, due to cancellations between the SM and the new scalar contributions, which eventually gets excluded by the false vacuum limits (black shaded). The near-degeneracy of $M_{T^\pm}$ and $M_{T^0}$ allows the DM bounds to be overlaid here, following the evaluation approach in ref. \cite{Bandyopadhyay:2024plc}. Firstly, the bright yellow points satisfy the observed relic  $\Omega_{obs} = 0.1198\pm 0.0012$ \cite{Planck:2018vyg}, starting from $M_{T^\pm}\sim 2.5$ TeV for $\lht=0$, and this mass bound increases further with higher $\abs{\lht}$. The black opaque region represents DM overabundance. Most of the remaining underabundant space however gets excluded by the LUX-ZEPLIN (LZ) limits (turquoise shaded) on direct detection of DM \cite{LZ:2024zvo}. The Sommerfeld-enhanced annihilation rates of the DM also strengthens indirect detection signals \cite{Katayose:2021mew}, with Fermi-LAT  \cite{MAGIC:2016xys} data excluding $M_{T^\pm} \sim 2-3.5$ TeV (purple shaded), and HESS (Einasto-2 profile) \cite{HESS:2022ygk} ruling out anything above $M_{T^\pm}\sim 1.6$ TeV (orange shaded), from the LZ-allowed space. Additionally, the compressed mass spectrum of this model leads to disappearing track (DT) signatures of $T^\pm$ at colliders, which excludes $M_{T^\pm} \lesssim 275$ GeV (yellow shaded), from a recast study \cite{Chiang:2020rcv} of the ATLAS limits \cite{ATLAS:2017oal}. Disheartening results emerge from these bounds, as the available parameter space is squeezed into an alley of $M_{T^\pm} \in [275 \text{ GeV}, 1.6 \text{ TeV}]$, with $0.02 \leq \abs{\lht}^{max} \leq 0.2$ corresponding to it, yielding at most $\sim$50\% of the observed DM relic, conclusively establishing ITM as insufficient for WIMP DM. This parameter space can be probed using DT \cite{Bandyopadhyay:2024plc}, or soft pion tracks \cite{Capdevilla:2024bwt} for lower triplet masses, arising from the $T^\pm \to T^0 \pi^\pm$ decay, favourably at a future muon collider.

	The absence of DM bounds yield a cleaner plot for HTM, as seen in \autoref{fig:exclusion}(b). The false vacuum, $\rgg$, and $\rzg$ exclusion regions remain similar to ITM, except for a dip around $M_{H^\pm} \sim 125$ GeV, caused by larger $\alpha_0$ affecting the $\rgg$ value. The tiny $v_t$ has negligible effect on both these measurements. Additional collider bounds, from the $2\tau2\nu$ final state search at LEP \cite{ALEPH:2013htx} (yellow shaded) and CMS \cite{CMS:2022syk} (cyan shaded) emanating from the $H^+ \to \bar{\tau}\nu_\tau$ decay, ruling out $M_{H^\pm} \lesssim$ 110 GeV, as pointed out in ref. \cite{Ashanujjaman:2024lnr}. In both \autoref{fig:exclusion}(a) and (b), the navy blue dashed line shows $\lht\simeq 3.5$, allowed for a perturbativity scale $\Lambda = 10^5$ GeV in the $Y=0$ scalar triplet models \cite{Bandyopadhyay:2025ilx}, which is our preferred upper limit.
	
	\section{Viability of FOPT}
	
	This ominous confinement of the $Y=0$ triplets have consequences in their possible contribution towards achieving a first order EWPT. For ITM, the convenience of a 2-step FOPT is lost with the demand of lower triplet masses and reasonably higher $\lht \gtrsim 1$ \cite{Patel:2012pi, Niemi:2020hto}, inevitably excluded by $\rgg,\rzg$ and LZ limits. With $ \mathcal{O}(1)\,\lht$ values, Ref. \cite{Bandyopadhyay:2021ipw} establishes an upper bound of $\mu_\mathcal{T} \lesssim 310$ GeV, i.e.  $M_{T^\pm} \lesssim 450$ GeV to have $\lht \lesssim 3.5$, that can still provide a 1-step FOPT considering only the doublet background field direction. The LZ bounds exclude this space as well, comprehensively incapacitating ITM for FOPT.
	
	Sacrificing the requirement for a WIMP DM and moving to HTM, a narrow band of the parameter space with FOPT possibilities exist within the $2\sigma$ allowed region of $\rgg$ and $\rzg$, shown in the inset of \autoref{fig:exclusion}(b).\footnote{The region aligns well with existing non-perturbative results for ITM \cite{Niemi:2020hto}, and for a complex scalar triplet \cite{Chala:2018opy}.} Here, the purple region can facilitate a 2-step FOPT with the correct choice of $\lht$ and $\lt$ combinations, while the orange region is only good enough for a 1-step FOPT. The calculations for the viability of the FOPT is performed with the thermal effective potential in terms of the background fields $\phi = \{\phi_h, \phi_t\}$ and temperature $T$:
	\begin{equation}\label{eq:Veff}
		V_{\rm eff}(\phi, T) = V_{0}(\phi) + V_{CW}(\phi) + V_{CT}(\phi) + V_{T}(\phi, T).
	\end{equation}
	
	Here, $V_0$ is the tree-level potential, $V_{CW}$ is the one-loop Coleman-Weinberg potential, $V_{CT}$ contains counterterms to enforce the tree-level EW VEV and mass values at one-loop, and $ V_{T}$ is the thermal correction, with additional daisy resummations, written in terms of field-dependent masses $m_i^2(\phi)$ wherever applicable, further details of which can be found in \appref{sec:app2}. We implement this potential in \texttt{CosmoTransitions} \cite{Wainwright:2011kj} to study the EWPT behaviour. Notably, the $V_{CW}$ depends on the renormalisation scale $Q$ via a logarithmic term $ \log\frac{m_i^2(\phi)}{Q^2}$, which influences our choice of benchmarks to illustrate the consequences of the 2-step FOPT. In the inset of \autoref{fig:exclusion}(b) we show the location of four benchmark points BP1-4, with $[M_{H^\pm},\lht]$ coordinates of [150 GeV, 1.00], [175 GeV, 1.38], [200 GeV, 1.7], and [250 GeV, 2.8], respectively. To avoid divergences in $V_{CW}$, we choose $Q$ =246 (340) GeV for BP1-3 (BP4). 
	
	\begin{figure}[h]
		\centering
		\subfigure[]{\includegraphics[width=0.35\linewidth]{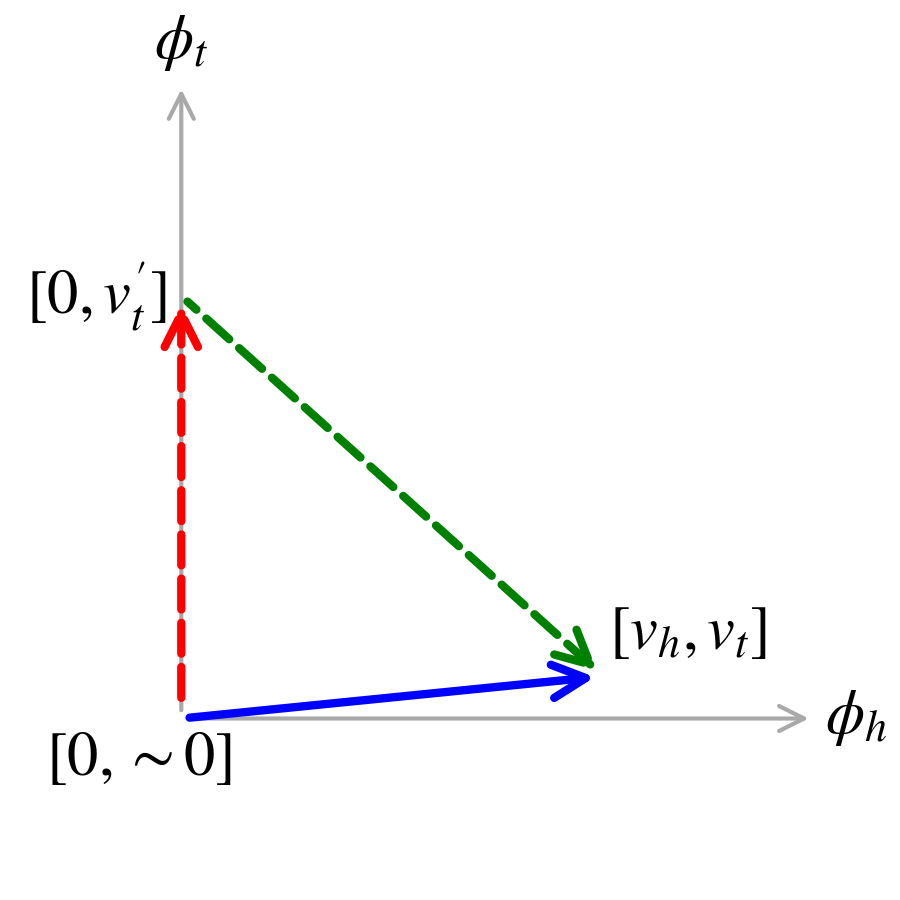}}
		\subfigure[]{\includegraphics[width=0.6\linewidth]{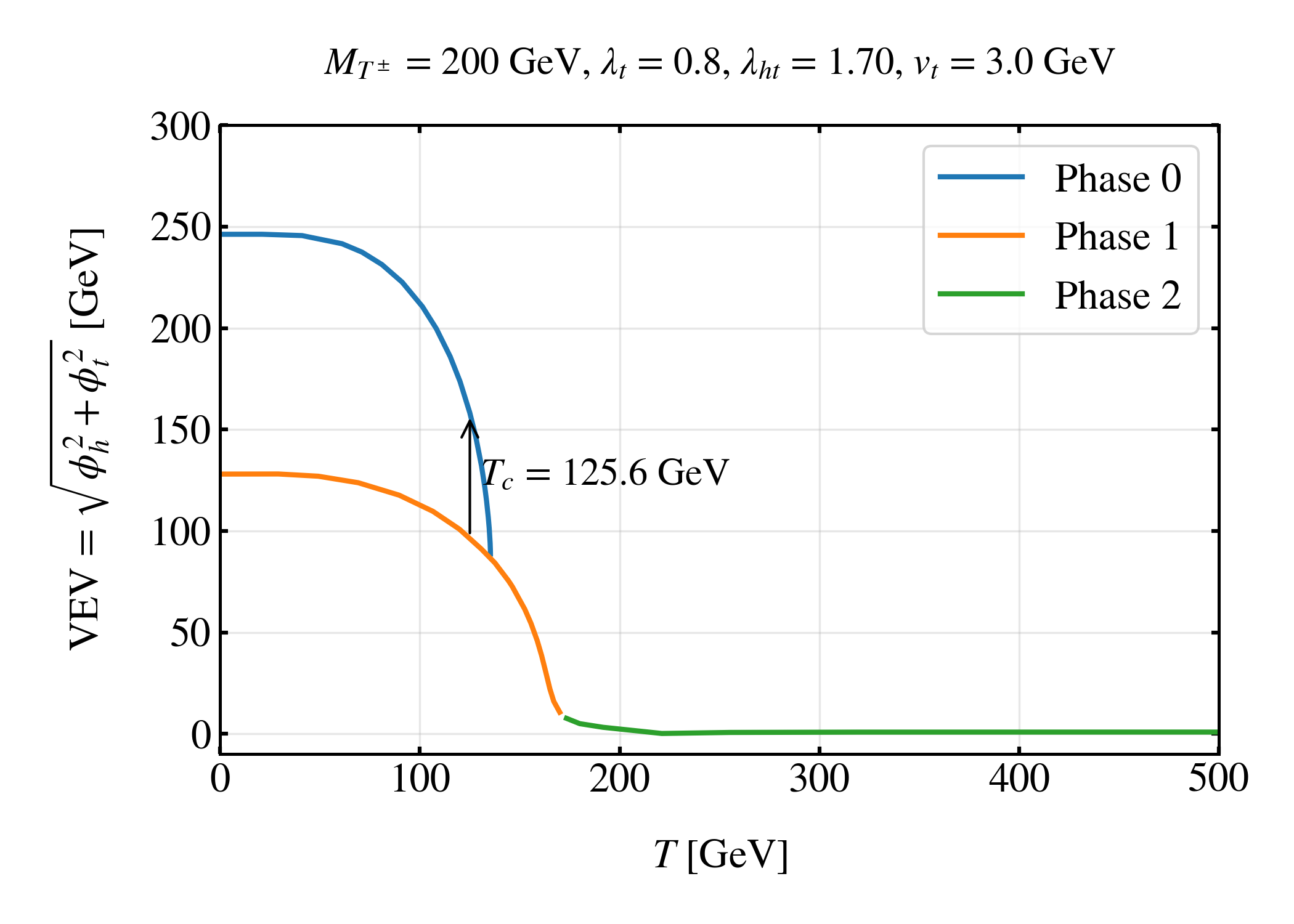}}
		\caption{(a) Schematics of a 1-step and 2-step PT. (b) 2-step FOPT phase diagram in BP3.}
		\label{fig:ptsteps}
	\end{figure}
	
	In \autoref{fig:ptsteps}(a) we show the schematics of a 2-step PT in context of HTM. It is noteworthy that, the presence of the $A_{ht}$-term in $V_{0}$ means that [0,0] is not the minima of $V_{eff}$ in the early universe, and hence we designate [0,$\sim$0] as the high-temperature minima. $[v_h,v_t]$ is the current EW minima at T=0, which can be reached by two ways. A direct transition from [0,$\sim$0]$\to[v_h,v_t]$ (blue solid arrow) is the 1-step PT, which can be second- or first-order. Otherwise, if condition (b) of \autoref{eq:truemin} is met, then at some finite temperature, the [0,$\sim$0] phase can transit to an intermediate $[0,v_t']$ phase (red dashed arrow), which can then tunnel to the $[v_h,v_t]$ phase via an FOPT (green dashed arrow). The FOPT is signalled at a critical temperature $T_c$ where two degenerate minima appear in the potential. \autoref{fig:ptsteps}(b) shows the transitions between the phases obtained in BP3, in terms of temperature vs the VEV $v=\sqrt{\phi_h^2 +\phi_t^2}$ of the phase. A smooth second-order transition takes place between phase 2 ([0,$\sim$0]) and phase 1 ($[0,v_t']$) at $T\sim 180$ GeV, while FOPT happens between phase 1 and phase 0 ($[v_h,v_t]$) at $T_c = 125.6$ GeV, which becomes the true EW minima at $T=0$ GeV. The $[0,v_t']$ phase is remnant as a shallow local minima \cite{Patel:2012pi}. For a 2-step PT, the strength of the transition is defined in terms of the difference in VEV between the two phases at $T=T_c$, with a strong FOPT implied by the criteria $\frac{\Delta v(T_c)}{T_c} \gtrsim 1$. This criteria can lead to EW baryogenesis with appropriate sources of CP violation.
	
	\section{Gravitational waves}
	
	For the aforementioned FOPT, the system tunnels across the potential barrier from the high-$T$ false vacuum decays into the true EW vacuum. The tunnelling begins at some temperature $T_n \lesssim T_c$, called the nucleation temperature of critical bubbles that determine the rate of the transitions. A criteria for successful FOPT demands $S/T_n \simeq 140$, where $S$ is the 3-d Euclidean action for the critical bubble \cite{Apreda:2001us}. The strength of the PT is then represented by the parameter $\alpha = \frac{\rho_{vac}}{\rho_{rad}({T_n})}$, where $\rho_{vac}$ is the vacuum energy density liberated during the PT, and $\rho_{rad}({T_n}) = g_* \pi^2 T_n^4/30$ is the total radiation energy density at $T=T_n$. The inverse time duration of the PT is estimated with the quantity $\beta/H_n = T_n \frac{dS}{dT}\Bigr|_{T_n}$, where $H_n$ is the Hubble rate at $T=T_n$ \cite{Chao:2017vrq}. These bubbles expand through the plasma with a bubble wall velocity estimated as $v_w = \frac{1/\sqrt{\alpha} + \sqrt{\alpha^2 + 2\alpha/3}}{1+\alpha}$ \cite{Kamionkowski:1993fg}. The collisions between these expanding bubbles lead to the stochastic production of GW at $T_n$, which can be detected at future space-based interferometers. The produced GW can be quantified in terms of the energy spectrum $\Omega_{\rm GW}h^2 (f)$ with $f$ being the frequency and $h\sim 0.7$, which can receive the following contributions \cite{Caprini:2015zlo}:
	\begin{equation}
		\Omega_{\rm GW}h^2 \simeq \Omega_{\rm col}h^2 + \Omega_{\rm sw}h^2 + \Omega_{\rm turb}h^2.
	\end{equation}
	
	Here, $ \Omega_{\rm col}$ sources from the collision of the bubble walls, encoding the contribution from the scalar fields themselves. $\Omega_{\rm sw}$ comes from the sound waves in the plasma after the collisions. $\Omega_{\rm turb}$, usually subdominant, arises from the magnetohydrodynamic turbulence in the plasma. Parametrising the GW production via FOPT in the set of parameters $\{T_n, \alpha, \beta/H_n\}$ computed with \texttt{CosmoTransitions}, we follow the prescription of ref. \cite{Caprini:2015zlo} to evaluate the components of the GW spectrum. 
	
	\begin{table}[h]
		\centering
		\renewcommand{\arraystretch}{1.2}
%		\begin{ruledtabular}
			\begin{tabular}{|c|c|c|c|c|c|c|}
				\hline
				BP & $\{M_{H^\pm} ({\rm GeV}), \lambda_{ht}, \lambda_t\}$ &\makecell{$T_c$ (GeV)}&$\frac{\Delta v(T_c)}{T_c}$&\makecell{$T_n$ (GeV)}&$\alpha$&$\beta/H_n$\\
				\hline
				BP1 & \{150, 1.00, 0.383\} &134.7&1.22&118.3&0.019&882.3\\
				BP2 & \{175, 1.38, 0.751\} &126.4&1.45&104.2&0.035&508.4\\
				BP3 & \{200, 1.70, 0.800\} &125.6&1.46&99.6&0.042&266.1\\
				BP4 & \{250, 2.80, 1.780\} &118.7&1.64&73.6&0.118&88.5\\
				\hline
			\end{tabular}
%		\end{ruledtabular}
		\caption{Benchmark points and corresponding FOPT parameters, evaluated with \texttt{CosmoTransitions}.}
		\label{tab:bp}
	\end{table}
	
	\begin{figure}[h]
		\centering
		\includegraphics[width=0.8\linewidth]{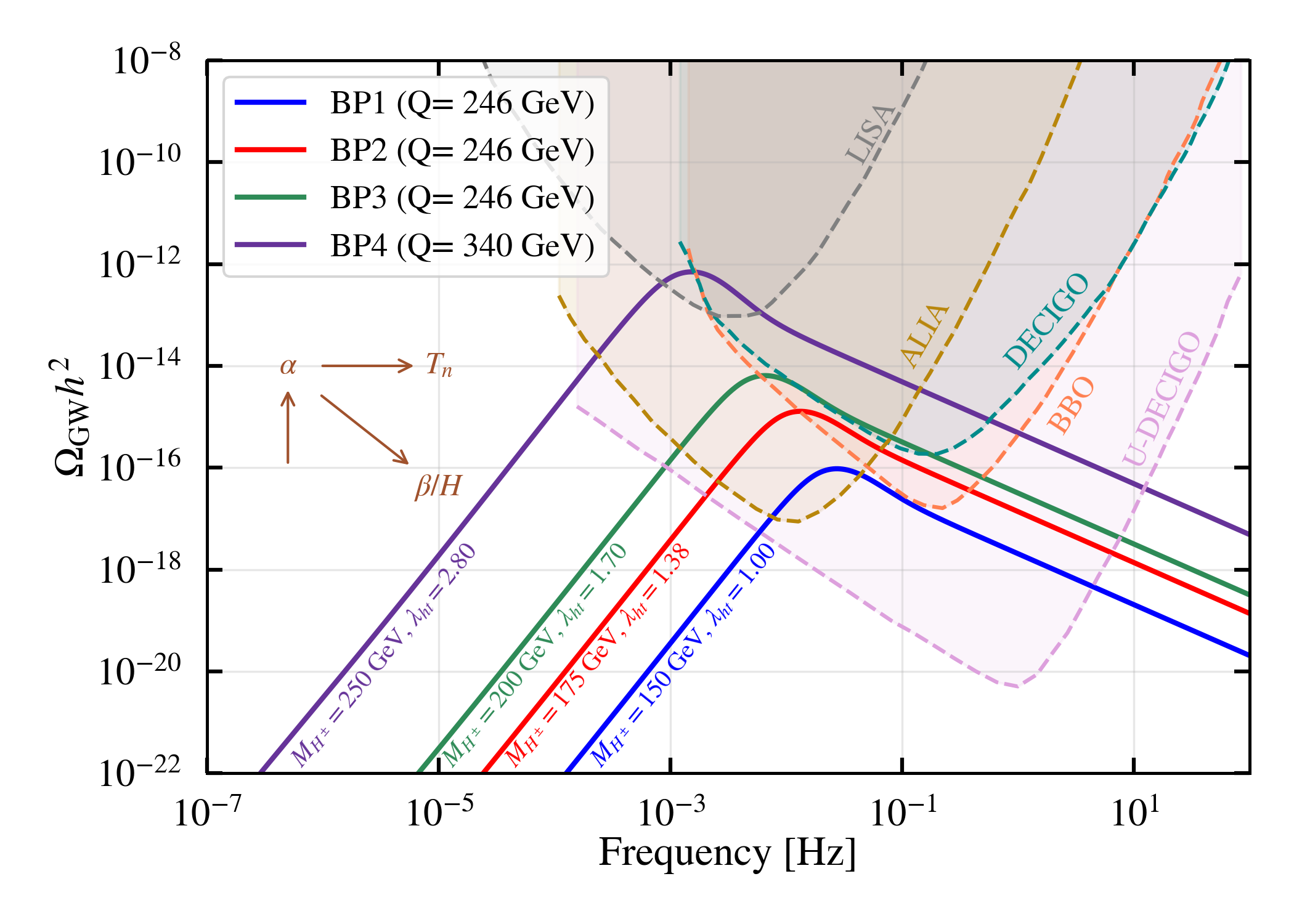}
		\caption{The GW spectra from the four BPs, overlaid with the PLISCs from various future interferometers.}
		\label{fig:gwspec}
	\end{figure}
	
	In \autoref{tab:bp} we display the FOPT parameters for BP1-4, keeping $v_t = 3$ GeV for all, as it has minimal impact on the parameters. Noticeably, an increase in $\lht$ enhances $\alpha$, and reduces both $\beta/H_n$ and $T_n$, which ideally results into better detectability at the GW interferometers. The subsequent GW spectra for the BPs are presented in \autoref{fig:gwspec}, with overlays of the power law-integrated sensitivity curves (PLISC) \cite{Thrane:2013oya} of five future space-based GW interferometers: LISA \cite{LISA:2017pwj}, ALIA \cite{Gong:2014mca}, BBO \cite{Corbin:2005ny}, DECIGO \cite{Kawamura:2020pcg}, and Ultimate-DECIGO (U-DECIGO) \cite{Kudoh:2005as}. The majority contribution in all BPs come from $ \Omega_{\rm col}$, with an extra bump at the peak from $\Omega_{\rm sw}$. Evidently, BP1 (blue line) with the weakest $\alpha$ yields the lowest peak $\Omega_{\rm GW}h^2$ at $\sim 10^{-16}$, detectable only at ALIA and U-DECIGO. Both BP2 (red line) and BP3 (green line) lie within the sensitivity range of ALIA, BBO, and U-DECIGO, with BP3 barely touching the DECIGO limit. The large $\alpha$ and small $\beta/H_n$ and $T_n$ enables a peak of $\sim 10^{-12}$ for BP4 (purple line), which can be sensed at LISA, alongside all the other detectors presented. The curves clearly showcase the impact of the three parameters in shifting the peaks, which we show with representative brown arrows in the plot.

	\section{Hints at colliders}
	
	The telltale sign of HTM is the presence of $v_t$ that violates the custodial symmetry at the tree level, leading to the $H^\pm Z W^\mp$ vertex $	g_{H^+ Z W^-} = (v_t g_2^2 \cos\theta_W \cos \alpha_+  - \frac{v_h}{2}g_1 g_2 \sin\theta_W \sin \alpha_+)$, which obviously vanishes as $v_t \to 0$. In the mass range relevant for a 2-step FOPT, this vertex can be directly probed at a lepton collider via the production channel $\ell^+ \ell^-  \to Z \to H^\pm W^\mp$, followed by the $H^\pm \to ZW^\pm$ decay, for which an $e^+ e^-$ collider like FCC-ee with maximum collision energy $\sqrt{s} = 350$ GeV seems ideal.  The primary irreducible background for this is from $\ell^+ \ell^-  \to VVV$ ($V\in \{W^\pm, Z\}$). A fully visible hadronic final state can enable the reconstruction of the triplet charged Higgs, but the large $t\bar{t}$ backgrounds at $\sqrt{s} = 350$ GeV implies the requirement of larger luminosities. Hence, for illustrative purposes, we resort to $\sqrt{s} = 280$ GeV, beneficial for masses upto BP3, owing to expectedly larger production rates and suppressed backgrounds.
	
	\begin{table}[h]
		\centering
		\renewcommand{\arraystretch}{1.2}
	
%		\begin{ruledtabular}
			\begin{tabular}{|c|p{1.5cm}|p{1.5cm}|p{1.5cm}|c|}
				\hline
				\multirow{2}{*}{$\sqrt{s}$ (GeV)} & \multicolumn{3}{c|}{\makecell{$\sigma_{H^\pm W^\mp}$ \\ $\times BR(H^\pm \to ZW^\mp)$ (fb)}} & \multirow{2}{*}{$\sigma_{VVV}$ (fb)} \\
				\cline{2-4}
				&\centering BP1&\centering BP2&\centering BP3&\\
				\hline
				280 & \centering 0.153 & \centering 0.110 & \centering 0.011 & 0.979\\
				\hline
			\end{tabular}
				\caption{Signal and background production rates at a 280 GeV leptonic collider for BP1-3.}
			\label{tab:csbp}
%		\end{ruledtabular}
	\end{table}
	
	\begin{figure}[h]
		\centering
		\includegraphics[width=0.9\linewidth]{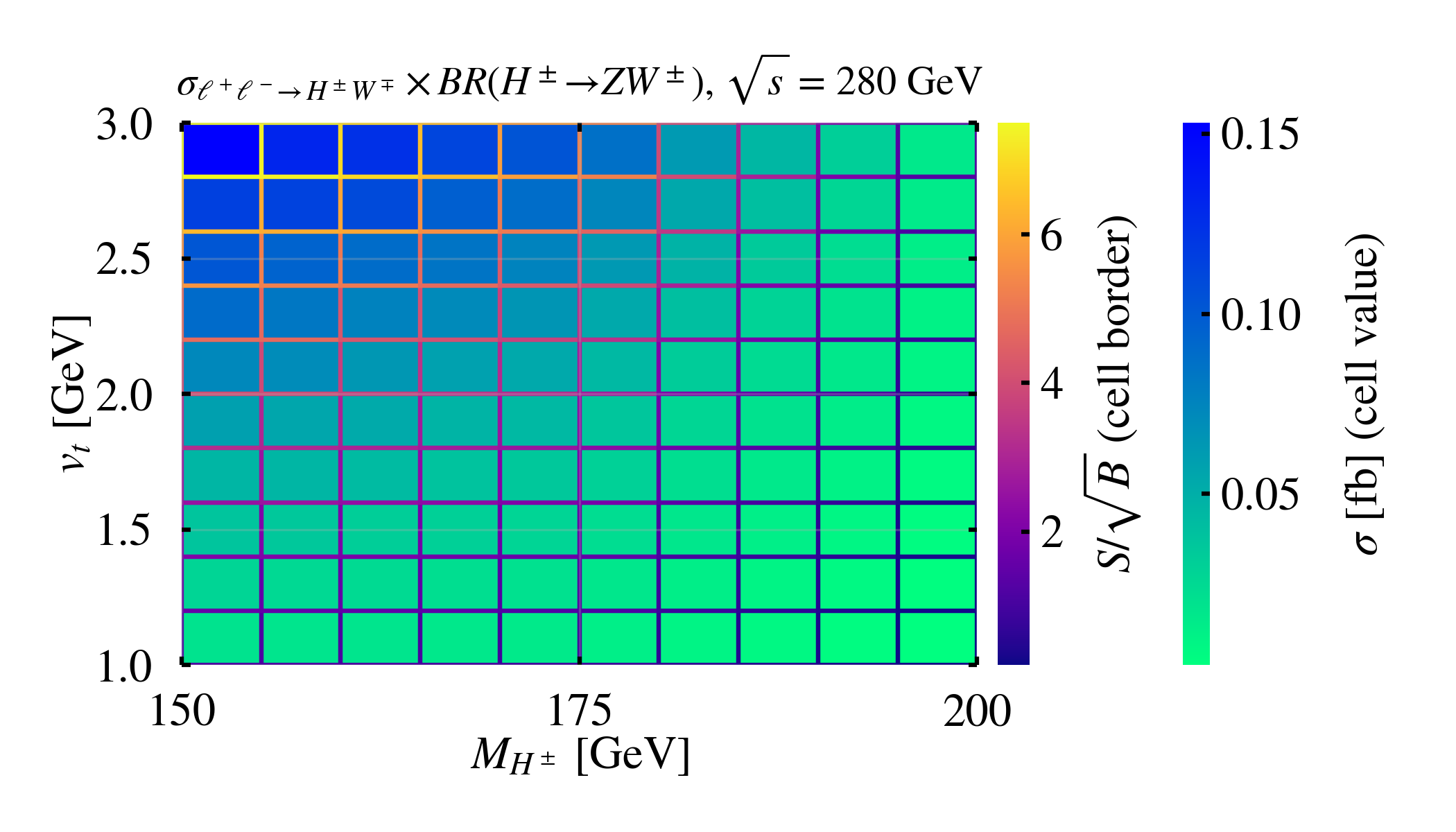}
		\caption{The cell values show $\sigma_{\ell^+ \ell^- \to H^\pm W^\mp} \times BR(H^\pm \to ZW^\mp)$ at $\sqrt{S}=$ 280 GeV, as a function of $M_{H^\pm}$ and $v_t$. The cell borders represent the $S/\sqrt{B}$ values assuming 60\% signal and 20\% background efficiencies.}
		\label{fig:csvt}
	\end{figure}
	
	\autoref{tab:csbp} presents the scaled signal rates $\sigma_{H^\pm W^\mp} \times BR(H^\pm \to ZW^\mp)$ for BP1-3 at a lepton collider with $\sqrt{s} = 280$, with the primary background rate $\sigma_{VVV}\sim1$ fb. The signal rates tend to be $\mathcal{O}(10^{-1}-10^{-2})$ times less than the background, which remain promising for BP1 and BP2.  While a detailed final-state analysis is beyond the scope of this letter, assuming a 60\% signal and 20\% background efficiency-- via cuts and/or boosted decision trees (BDT) \cite{Bandyopadhyay:2024gyg}-- in the fully visible hadronic final state, the projected $S/\sqrt{B}$ significances are shown as a function of $M_{H^\pm}, v_t$, and the scaled signal rate, in \autoref{fig:csvt}. Here, the cell colours represent the $\sigma\times BR$, while the cell borders represent $S/\sqrt{B}$, at a target luminosity of 5 \abi. We see that, $\gtrsim 3\sigma$ sensitivity of $v_t \simeq 3$ GeV is possible for $M_{H^\pm} \lesssim 190$ GeV, and $\gtrsim5\sigma$ discovery of the triplet nature is viable for $v_t \gtrsim 2$ GeV, when $M_{H^\pm} \simeq 150$ GeV. For BP1-3, the significances are be $\sim 7.8\sigma, \sim 5.6\sigma, \sim 1\sigma$ respectively.

	At the 13.6 TeV LHC, the $pp\to H^\pm W^\mp$ production rates are $\sim5$ times more than at the lepton collider, but the dominant subprocess $gg \to h \to H^\pm W^\mp$ is insensitive to $v_t$. $pp\to W^\pm \to H^\pm Z$ is an alternative with similar cross-sections, but is likely to suffer from the overwhelming $\sim 1$ pb rates of $pp\to VVV$ \cite{CMS:2020hjs}. Ref. \cite{Ashanujjaman:2024lnr} also elaborates on the viability of the allowed mass range up to 200 GeV in LHC searches from $pp \to H^\pm H^0$, while 200-500 GeV charged triplet mass is explored in ref. \cite{Bandyopadhyay:2024gyg} at a muon collider, for a singlet DM-extended HTM. LHC  phenomenology of  supersymmetric triplet \cite{Bandyopadhyay:2014vma, Bandyopadhyay:2015ifm} and muon collider phenomenology of  complex triplet \cite{Bandyopadhyay:2020otm} also explore similar production modes.

	\section{Conclusions}	
	
 Our perusal into the broad phenomenological appeal of hyperchargeless scalar triplets have thus revealed the disheartening requirement of a trade-off between having a DM candidate and the viability of an FOPT. In reality, the most recent DM limits render the $Z_2$-odd ITM to be insufficient for the observed DM relic, and discards any chance of achieving the FOPT, confining the parameter space to $M_{T^\pm} \in [275 \text{ GeV}, 1.6 \text{ TeV}]$, with $0.02 \leq \abs{\lht}^{max} \leq 0.2$. Dropping the $Z_2$ symmetry and the DM hope altogether, the non-inert HTM presents a narrow space between $M_{H^\pm} \in [150 \text{ GeV}, 275 \text{ GeV}]$ corresponding to $1.0 \leq \lht^{max} \leq 3.5$, where the sought-after 2-step FOPT remains feasible, leading to stochastic GW detectable at future interferometers. Avenues for probing the custodial-symmetry breaking nature of this triplet by looking for $v_t$-dependent charged scalar production, in the same region, are within reach of a high-luminosity lepton collider. The trade-off can be ameliorated by extending HTM with an inert singlet DM such as in \cite{Bandyopadhyay:2024gyg}, a detailed inverstigation of FOPT patterns and EW baryogenesis of which is deferred to a future work.
 
 \subsection*{Acknowledgement}
	
SP acknowledges the Council of Scientific and Industrial Research (CSIR), India for funding his research (File no: 09/1001(0082)/2020-EMR-I).

	\appendix

	\section{Functions for $h\to\gamma\gamma$ and $h\to Z\gamma$} \label{sec:app1}
	
	\begin{figure}[h]
		\centering
		\subfigure[]{\includegraphics[width=0.32\linewidth]{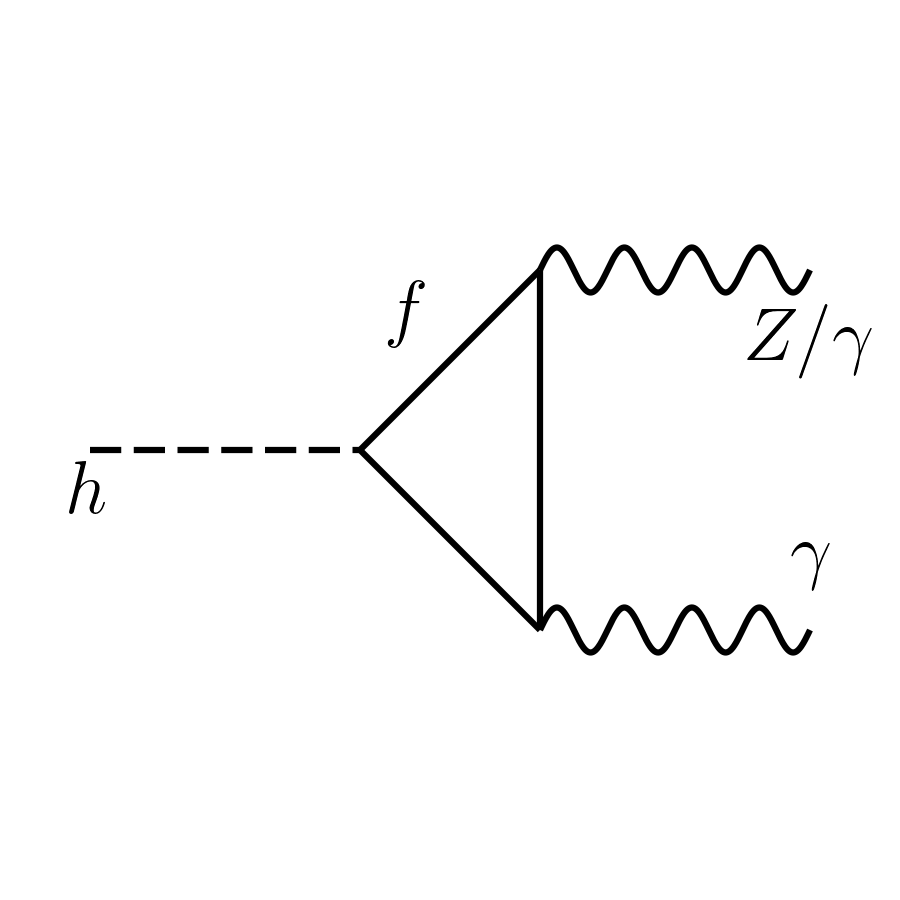}}
		\subfigure[]{\includegraphics[width=0.32\linewidth]{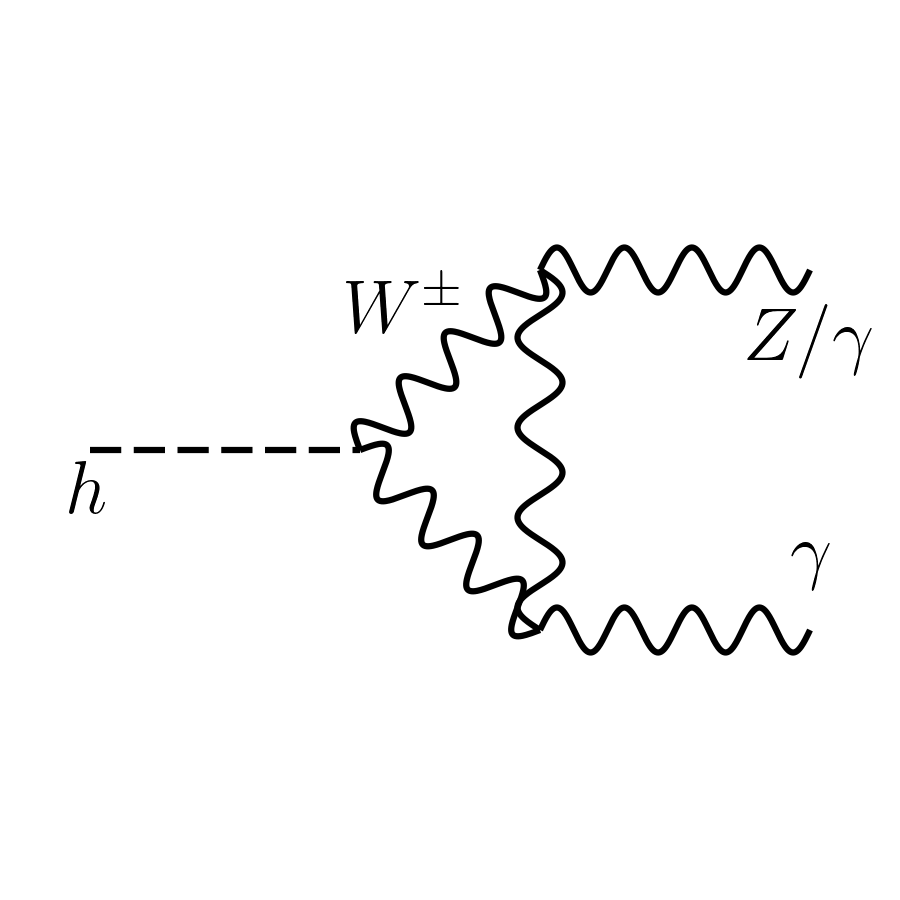}}
		\subfigure[]{\includegraphics[width=0.32\linewidth]{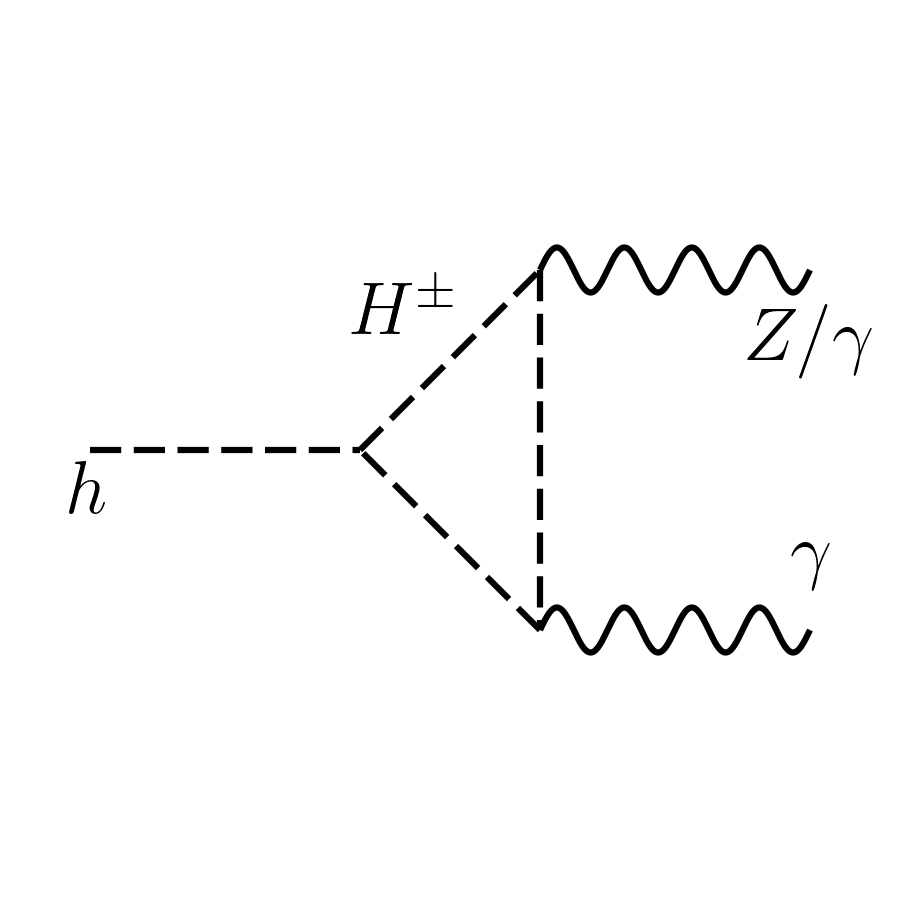}}
		\subfigure[]{\includegraphics[width=0.32\linewidth]{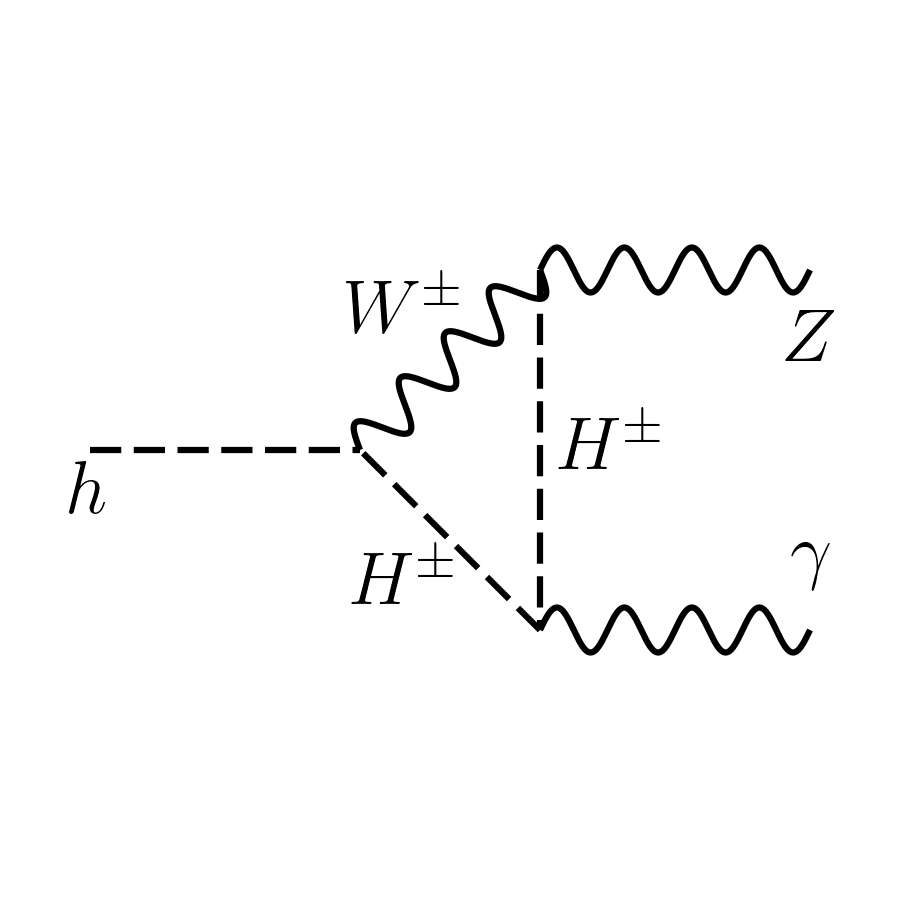}}
		\subfigure[]{\includegraphics[width=0.32\linewidth]{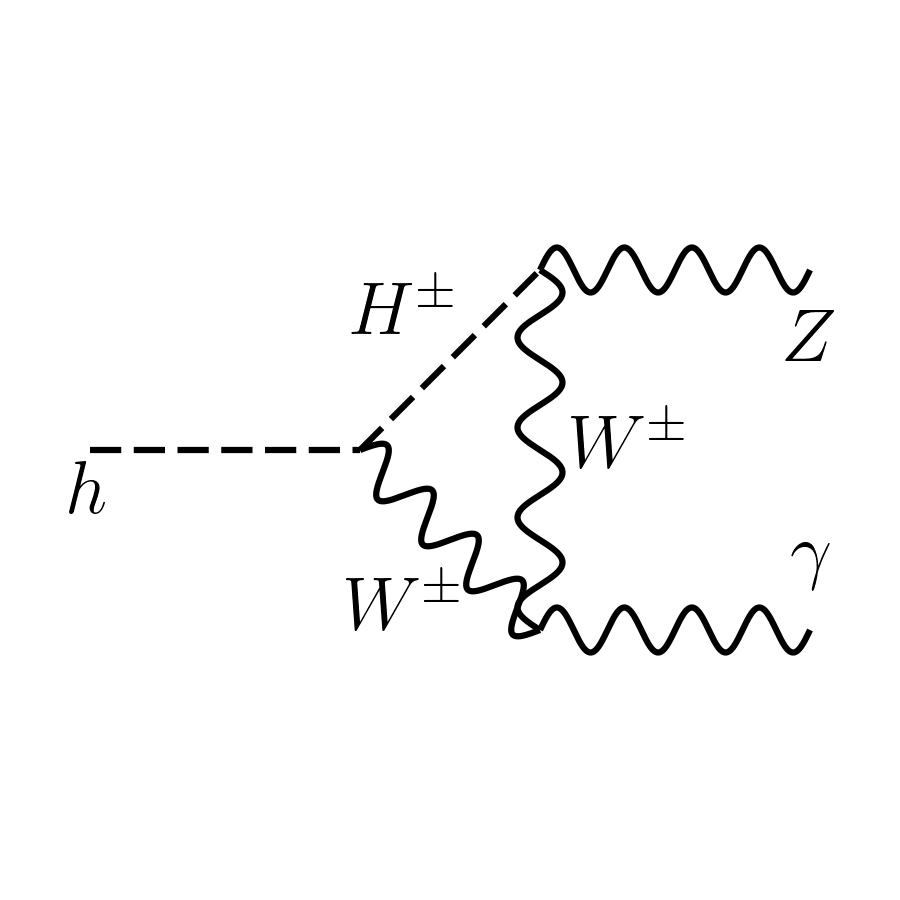}}
		\caption{Feynman diagrams contributing to the $h\to Z\gamma/\gamma\gamma$ decay in HTM: (a) fermion loop, (b) $W$-boson loop, (c) charged scalar loop, (d) VSS loop, (e) VVS loop.}
		\label{fig:fdloop}
	\end{figure}
	
	For $h\to\gamma\gamma$ in HTM, only the fermion loop, $W$-boson loop, and the charged scalar loop in \autoref{fig:fdloop}(a)-(c) contribute to the decay. The expression for the partial decay width reads \cite{Djouadi:2005gi,Djouadi:2005gj}:
	\begin{align}
		\begin{split}
			\Gamma_{h\to\gamma\gamma} =&{} \frac{G_F \alpha_{EM}^2 M_h^3}{128\sqrt{2}\pi^3}  \Bigr| \sum_{f}N_f^c  Q_f^2 \mathcal{A}_f^{\gamma\gamma}(X_{M_f}) g_{hff}  \\ &{} +  \mathcal{A}_W^{\gamma\gamma}(X_{M_W}) g_{hWW} + \frac{c_{hH^+H^-}}{2M_{H^\pm}^2} \mathcal{A}_{S}^{\gamma\gamma}(X_{M_{H^\pm}}) \Bigr|^2
		\end{split}
	\end{align}
	
	Here, $G_F$ is the Fermi constant, $\alpha_{EW} = e^2/4\pi$ is the EW coupling constant, $N_f^c$ and $Q_f$ represent the colour factor and EM charge of the fermions, $X_{M_i} = \frac{M_h^2}{4M_i^2}$, $g_{hff} = \cos\alpha_0$, $g_{hWW} = \cos\alpha_0 + 4(v_t/v) \sin\alpha_0$ , and $c_{hH^+H^-} \approxeq \cos\alpha_0 \lht v^2$. The loop functions $\mathcal{A}_i^{\gamma\gamma}$ are given by:
	\begin{align}
		&{} \mathcal{A}_f^{\gamma\gamma}(x) = 2\left[x+(x-1)f(x)\right]x^{-2},\nonumber \\
		&{} \mathcal{A}_W^{\gamma\gamma}(x) = -\left[2x^2+3x+3(2x-1)f(x)\right]x^{-2},\\
		&{} \mathcal{A}_S^{\gamma\gamma}(x) = -\left[x-f(x)\right]x^{-2},\nonumber
	\end{align}
	where we have:
	\begin{equation}
		f(x) = \begin{cases} 
			\sin^{-1}\sqrt{x} & x\leq 1 \\
			-\quart \left[\log\frac{1+\sqrt{1-1/x}}{1-\sqrt{1-1/x}} -i\pi \right] ^2& x>1. \\
		\end{cases}
	\end{equation}
	
	The partial decay width expression for $h\to Z\gamma$ obtains contributions not only from the fermion, $W$-boson and charged scalar loops, but also the vector-scalar-scalar (VSS) and vector-vector-scalar (VVS) loops shown in \autoref{fig:fdloop}, given as follows \cite{Degrande:2017naf,Hue:2017cph}:
	
	\begin{align}
		\begin{split}
			\Gamma_{h\to Z\gamma} ={}& \frac{M_h^3}{32\pi} \left(1-\frac{M_Z^2}{M_h^2}\right)^3 \Biggr| \frac{-\alpha_{EW}}{2\pi v} \\ & \Biggr(\sum_{f} \frac{N_f^c Q_f V_f}{\cos\theta_W} \mathcal{A}_{f}^{Z\gamma}(\tau_{M_f},\lambda_{M_f}) g_{hff} \\ & + \mathcal{A}_W^{Z\gamma}(\tau_{M_W},\lambda_{M_W}) g_{hWW} \\ & +  \frac{c_{hH^+H^-}}{2M_{H^\pm}^2} \mathcal{A}_S^{Z\gamma}(\tau_{M_{H^\pm}},\lambda_{M_{H^\pm}}) \Biggr) \\ & + 2(\mathcal{A}_{VSS} + \mathcal{A}_{VVS})\Biggr|^2.
		\end{split}
	\end{align}
	
	Here, $V_f = 2(T_3^f - 2Q_f  \sin^2\theta_W),\, \tau_{M_i} = \frac{4M_i^2}{M_h^2}, \, \lambda_{M_i} = \frac{4M_i^2}{M_Z^2}$, and the various loop functions are given by:
	\begin{align}\label{eq:Azg}
		\begin{split}
			&{} \mathcal{A}_{f}^{Z\gamma}(x,y) = \frac{-1}{\sin\theta_W} (\mathcal{I}_1 (x,y) - \mathcal{I}_2 (x,y)), \\
			&{} \mathcal{A}_{W}^{Z\gamma}(x,y) = \frac{-\cos\theta_W}{\sin\theta_W}\Biggr( 4(3-\tan^2\theta_W)\mathcal{I}_2(x,y)  \\  &\quad + \left((1+2x^{-1})\tan^2\theta_W -(5+2x^{-1}) \right)\mathcal{I}_1(x,y) \Biggr),\\
			&{} \mathcal{A}_{S}^{Z\gamma}(x,y)  = 2c_{ZH^+H^-}\mathcal{I}_1(x,y),\\
			&{} \mathcal{A}_{VSS} = c_{H^\pm Z W^\mp} c_{H^\pm h W^\mp} \frac{-\alpha_{EW}^2}{M_W^2}\int_{0}^{1}dx\int_{0}^{1}dz \frac{\mathcal{I}_{VSS}}{\Delta_{VSS}},\\
			&{} \mathcal{A}_{VVS} =  c_{H^\pm Z W^\mp} c_{H^\pm h W^\mp} \frac{-\alpha_{EW}^2}{2M_W^2}\int_{0}^{1}dx\int_{0}^{1}dz \frac{\mathcal{I}_{VVS}}{\Delta_{VVS}}.
		\end{split}
	\end{align}
	
	In the above expressions, the couplings $c_{ijk}$ are $c_{ZH^+H^-} = \frac{1-2\sin^2\theta_W}{\sin\theta_W\cos\theta_W}$, $c_{H^\pm Z W^\mp} = \frac{1}{2\sin^2\theta_W}$ $\left( -v_h \frac{\sin^2\theta_W}{\cos\theta_W}\sin\alpha_+ + 2v_t \cos\theta_W \cos\alpha_+ \right)$, and $c_{H^\pm h W^\mp} = \frac{-1}{\sin\theta_W}\left( \half \cos\alpha_0\sin\alpha_+ + \sin\alpha_0\cos\alpha_+ \right)$. The additional functions $\mathcal{I}_{VSS,VVS}$ and $\Delta_{VVS,VSS}$ for the VSS and VVS loops are the same ones in Equations 20 and 23 of ref. \cite{Degrande:2017naf}. Notably, the contributions from the VVS and VSS loops are negligible for a model like HTM where $v_t \ll v_h$. For ITM, only \autoref{fig:fdloop}(a)-(c) contribute for both $h\to\gamma\gamma/Z\gamma$, the expressions for which can be obtained by putting $\alpha_0 = \alpha_+ = 0$ wherever applicable, and $c_{ZT^+T^-} = \cos\theta_W/\sin\theta_W$ in $\mathcal{A}_S^{Z\gamma}(x,y)$. The definition of the signal strengths for a model $M\in$\{ITM, HTM\} can be simplified into:
	\begin{equation}
		\mu_{ij}^M = (\cos^2{\alpha_0})^M \times \frac{\Gamma_{h\to ij}^{M} \times \Gamma_{\hsm,tot}}{\Gamma_{\hsm\to ij} \times \Gamma_{h,tot}^{M}}, \label{eq:Rij}
	\end{equation}
	where for ITM we have $(\cos^2{\alpha_0})^{\rm ITM} =  1$, and $h \to \hsm$.
	%\begin{align}
	%	\begin{split}
		%		&{} \Delta_{VSS} = -x(1-x) M_Z^2 + (1-x) M_{W}^2 + x M_{H^\pm}^2 \\ 
		%		&\quad - xz(1-x)(m_h^2 - M_Z^2), \\
		%		&{} \mathcal{I}_{VSS} = x^2 z \left[ 
		%		\frac{2}{3} x^2(1+z) + \frac{2}{3} x(1 - 2z) - 1 \right] m_h^2 \\ 
		%		&\quad+ x^2 (z - 1) \left[ \frac{2}{3} x(2 - x)(z - 2) + 1 \right] M_Z^2 \\
		%		&\quad+ x \left[ \frac{2}{3}(z+1)x^2 - 5x + 4 \right] M_W^2 \\
		%		&\quad + x^2 \left[ -\frac{2}{3}(1+z)x + 1 \right] M_{H^\pm}^2,\\
		%		&{} \Delta_{VVS} =  -x(1-x) M_Z^2 + (1-x) M_{H^\pm}^2 + x M_{W}^2 \\ 
		%		&\quad - xz(1-x)(m_h^2 - M_Z^2), \\
		%		&{} \mathcal{I}_{VVS} = \left[ -\frac{2}{3} x^4 z(1 + z) + \frac{8}{3} x^3 z(2 - z) - 3x^2 z \right] M_h^2 \\ &\quad + \Bigg[ 
		%		\frac{2}{3} x^4(z - 2)(z - 1) + \frac{4}{3} x^3(2z^2 - 3z + 1) \\ &\quad+ x^2(z - 1) \Bigg] M_Z^2 + \left[ \frac{-2}{3}x^3(1+4z) +x^2(6z-1) \right] M_{H^\pm}^2 \\ &\quad
		%		+ \left[ \frac{2}{3}x^3(4z+1)+3x^2(3-2z) \right] M_W^2.
		%	\end{split}
	%\end{align}

	\section{Thermal effective potential for HTM} \label{sec:app2}
	
	In \autoref{eq:Veff}, first component of $V_{eff}(\phi,T)$ is the tree-level potential,  in terms of the background fields $\phi_h, \phi_t$:
	\begin{equation}
		V_{0} =  \frac{ \mu_\Phi^2 \phi_h^2 + \mu_\mathcal{T}^2 \phi_t^2}{2}   + \frac{\lh\phi_h^4 +\lt\phi_t^4 + \lht\phi_h^2 \phi_t^2 -A_{ht}\phi_h^2 \phi_t}{4}.
	\end{equation}
	
	The Coleman-Weinberg term involves the field-dependent masses $m_i(\phi)$ and the renormalisation scale $Q$, as follows:
	\begin{equation}
		V_{CW} = \frac{1}{64 \pi^2} \sum_{i} n_i m_i^4(\phi) \left( \log\frac{m_i^2(\phi)}{Q^2} - c_i \right),
	\end{equation}
	
	Here, we have $i \in \{ W,Z,h,H^0,H^\pm,G^{0,\pm},t \}$, $n_i$ are the corresponding degrees of freedom (dof), $c_i$ are predefined constants for the CW potential, and $Q$ is the renormalisation scale. The dof values are $n_W = 6, n_Z = 3, n_t = -12, n_{h,H^0,G^0} = 1, n_{H^\pm, G^\pm} = 2$. For transverse vector bosons, $c_i = 1/2$, while for scalars, fermions, and longitudinal vector bosons, $c_i = 3/2$. To negate the shifts in the EW VEV and zero-temperature masses caused by $V_{CW}$, we add the counterterm potential:
	\begin{equation}
		V_{CT} = \delta \mu_\Phi^2 \phi_h^2 + \delta  \mu_\mathcal{T}^2 \phi_t^2 + \delta \lh\phi_h^4 +\delta\lt\phi_t^4 + \delta\lht\phi_h^2 \phi_t^2,
	\end{equation}
	
	the terms of which can be evaluated by solving the on-shell renormalisation conditions:
	\begin{equation}
		\frac{\partial(V_{CW} + V_{CT})}{\partial \phi_1} \Bigr|_{[v_h,v_t]}= 0, \frac{\partial^2(V_{CW} + V_{CT})}{\partial \phi_i \phi_j} \Bigr|_{[v_h,v_t]} = 0.
	\end{equation}
	
	In this approach, the Goldstone masses need to be regularised as $m_{G_i}^2(\phi) \to m_{G_i}^2(\phi) + \mu_{IR}^2$, where we keep $\mu_{IR} = 1$ GeV \cite{Ghosh:2022fzp}. Next, the finite temperature corrections are given by the thermal potential component:
	\begin{equation}
		V_{T} = \frac{T^4}{2\pi} \left[ \sum_{i} n_i \int_0^\infty a^2 \log(1\mp e^{\sqrt{a^2 + \frac{m_i^2(\phi)}{T^2}}})  \right].
	\end{equation}

	In $V_{T}$, the $\mp$ correspond to bosons and fermions, respectively. The consistency of thermal contributions require the resummation of daisy diagrams in the effective potential. Two primary methods, known as the Parwani \cite{Parwani:1991gq} and Arnold-Espinosa \cite{Arnold:1992rz} methods, exist for the daisy resummation. We follow the Parwani prescription, where the field-dependent masses are shifted to the thermal Debye masses by $m_i^2(\phi) \to m_i^2(\phi) +\Pi(T^2)$, with $\Pi$ being the daisy coefficients, evaluated as:
	\begin{equation}
		\Pi = C_{ij}T^2 = \frac{\partial^2 V_T}{\partial \phi_i \phi_j}\Bigr|_{T^2 \gg m^2}
	\end{equation}
	
	Only the scalars and the longitudinal components of the gauge bosons receive the thermal corrections \cite{Bandyopadhyay:2021ipw}. For the doublet and triplet scalars, the daisy coefficients are of interest:
	\begin{align}
		&{} C_h = \frac{1}{16}(3g_2^2 + g_1^2 + 4Y_t^2 + 8\lh + 2\lht),\\
		&{} C_t = \frac{1}{12}(6g_2^2 + 5\lt + 2\lht).
	\end{align}
	
	These are useful when one wants to study a minimal version of the FOPT behaviour using the high-temperature approximation \cite{Patel:2012pi}.
	
	The field-dependent masses of the neutral CP-even scalars are given by the eigenvalues of the mass matrix:
	\begin{equation}
		\mathcal{M}_0 = \begin{pmatrix}
			\mu_\Phi^2 + 3\lh\phi_h^2 +\frac{\lht}{2}\phi_t^2 -\frac{A_{ht}}{2} \phi_t & \phi_h(\lht\phi_t-\frac{A_{ht}}{2})\\
			\phi_h(\lht\phi_t-\frac{A_{ht}}{2}) &  \mu_\mathcal{T}^2 + 3\lt\phi_t^2 + \frac{\lht}{2}\phi_h^2 \\
		\end{pmatrix}
	\end{equation}
	
	The same for the charged triplet scalar $H^\pm$ and the Goldstone bosons $G^\pm$ are given by the eigenvalues of:
	\begin{equation}
		\mathcal{M}_\pm = \begin{pmatrix}
			\mu_\Phi^2 + \lh\phi_h^2 +\frac{\lht}{2}\phi_t^2 +\frac{A_{ht}}{2} \phi_t & \frac{A_{ht}}{2}\phi_h\\
			\frac{A_{ht}}{2}\phi_h &  \mu_\mathcal{T}^2 + \lt\phi_t^2 + \frac{\lht}{2}\phi_h^2 \\
		\end{pmatrix}
	\end{equation}
	
	The rest of the field-dependent masses are as follows:
	\begin{align}
		&{} m_{G^0}^2 = \mu_\Phi^2 + \lh\phi_h^2 +\frac{\lht}{2}\phi_t^2 -\frac{A_{ht}}{2} \phi_t, \\
		&{} m_{W}^2 = \frac{g_2^2}{4}(\phi_h^2 + 4\phi_t^2), \, m_{Z}^2 = \frac{g_1^2 + g_2^2}{4}\phi_h^2,\, m_t^2 = \frac{Y_t^2}{2}\phi_h^2.
	\end{align}
	
	\vspace{3em}
	
	\bibliography{TZgamma_refs}
	
\end{document}